\documentclass[12pt,twoside]{article}
\usepackage{fleqn,espcrc1}
\usepackage{graphicx}

\newcommand{\AmS}{{\protect\the\textfont2
  A\kern-.1667em\lower.5ex\hbox{M}\kern-.125emS}}
\newcounter{saveeqn}%

\title{Pion production from a critical QCD phase}

\author{N.G. Antoniou, Y.F. Contoyiannis, F.K. Diakonos,\\
A.I. Karanikas and C.N. Ktorides
\address{Department of Physics, University of Athens, GR-15771 Athens,
Greece}}

\begin{document}

\maketitle
\begin{abstract}
A theoretical scheme which relates
multiparticle states generated in ultrarelativistic nuclear
collisions to a QCD phase transition is considered
in the framework of the universality class provided by the 3-D Ising
model. Two different evolution scenarios for the QGP system are
examined. The statistical mechanics of the critical state is accounted
for in terms of (critical) cluster formation consistent with suitably cast 
effective action functionals, one for each considered type of expansion.
Fractal properties associated with these clusters,
characterizing the density fluctuations near the QCD critical point, 
are determined. Monte-Carlo simulations
are employed to generate `events', pertaining to the total system, which
correspond to signals associated with unconventional sources of pion
production. 

\end{abstract}

\section{Introduction}

It is widely speculated that ultra-relativistic nucleus-nucleus (A+A)
collisions offer an experimental methodology by which a transition from
ordinary (confined) matter to a quark-gluon plasma (QGP) state can be
transiently attained. In view of such a prospective it becomes imperative
to identify specific experimental signals in the outcome of these collisions
whose origin can be unmistakenly ascribed to a critical behaviour.

From a phenomenological point of view we can list three basic topics as
the most relevant to such an investigation:

(a) How to pick up patterns in the multihadron (pion) states generated in
A+A collisions from the central rapidity region and extract pertinent
information from them.

(b) What is one's conception of the evolution picture taking place between
the (presumed) formation of the quark-gluon plasma system and
hadronization.

(c) In what way does the underlying microscopic theory, i.e., QCD, leave its
imprint on the production patterns of hadrons.

There is a variety of attitudes that can be taken with respect to the above
issues. Even though the present work focuses upon the last one, it is
important to specify the adopted viewpoints on the other two as well.

In regard to (a), our premise draws its inspiration from
the Bialas-Peschanski \cite{BiaPe}
approach to data analysis for large multiplicity
events. The emphasis, here, lies on the possible presence of 
non-statistical deviations which can be uncovered from
experimentally determined probability distributions. The idea is
to look for big bursts from small regions (cells) of the phase space 
which signal unusual
fluctuations of density, an occurence that has been termed as search for
intermittency. Such intermittent spiky events induce a dependence of the
scaled moments of experimentally defined probability distribution
$P(p_1,...,p_n)$
on the size $\delta \Omega $ of each cell, 
according to some power-law. Searches for intermittency in multiparticle
states resulting from A+A collisions have so far revealed only
fluctuations due to Bose-Einstein interference.
However, the question that still remains open is whether intermittency
could be directly ascribed to a quark-hadron phase transition of second
order. This is precisely the attitude that underlines
our present approach, {\it i.e.} our intention is to propose a scheme
according to which intermittency patterns in a multiparticle hadronic
state are intimately linked with fluctuations which make their
presence near the QCD critical point.

Concerning, next, the space-time evolution of the (plasma) system produced
in the central rapidity region we shall adopt, for the most part, the
picture
promoted by Bjorken \cite{Bjork,Land}.
The basic idea is that, following the collision, a
``central plateau'' is formed comprised of a thin slab which contains quarks
and gluons in (local) thermal equilibrium.
The expansion is envisioned to be smooth
enough so that
conditions of thermal equilibrium persist on a locus which is monitored by
a collection of inertial observers with relative velocities. A proper
time ($\tau $) dependence of all relevant thermodynamic quantities that
describe the ``fluid'' emerges so that one writes $\varepsilon \left( \tau
\right)$,  $p\left( \tau \right)$, 
$\beta \left( \tau \right) $, respectively for energy and pressure densities
and for inverse temperature. Dynamical evolution is thereby described by a
sequence of a hyperboloidal, spacelike surfaces whose profile is specified
by a hyperbolic section in the $t-z$ plane ($z$ is the direction along which
the collision takes place) and flat geometry in the transverse directions.
This situation persists for a time scale of $\sim $ 10-20 fm/c $\left( \sim
\beta _cA^{\frac 13}\right) $ at which point formation of primordial 
hadrons (mostly pions) occurs. As an alternative possibility to the above
we shall consider, in Section 5, a spherically symmetric evolution of a
centrally formed fireball of quarks and gluons. Our numerical studies will
cover both scenarios. 

The final issue, towards which the present paper basically addresses
itself, concerns the interpolation between the underlying
microscopic theory and observed patterns in the multihadron states
generated by A+A collisions. In particular, we shall focus our efforts on 
determining possible imprints left by long range fluctuations, present
in a
critical quark-gluon system, on the produced multiparticle states.
Evidence is, in fact, mounting from ongoing theoretical activity based on
microscopic considerations which simulate QCD through specific models
({\it e.g.} that of Nambu and Jona-Lasinio), which lends
overwhelming support to the existence of a critical point in the
thermodynamical plane of temperature vs. chemical potential
\cite{Others,BerRaj}.
The latter is the remnant of a tricritical point occuring in the 
phase diagram of the ideal case $m_u=m_d=0$ and from which a second order
chiral phase transition curve originates belonging to the universality 
class of the O(4) $\sigma$-model \cite{O4}. For the realistic case where
the up-down current quark masses are different from zero it  
marks the termination of a first order phase transition curve
\cite{PhasDia}.
Such a point signals the onset of long range correlations and
the relevant issue is to
identify its universality class. It is widely believed
that the latter is represented by the 3-D Ising model \cite{BerRaj,3DIs}.
Under this
assumption we shall adopt a
strategy according to which the corresponding effective action  $\Gamma
\left[ \sigma \right]$ is: a) accomodated on
hyperboloidal surfaces of the type
specified by Bjorken's picture for the space-time expansion of
the plasma or b) adjusted to a spherical expansion scenario.
Given that $\Gamma \left[ \sigma \right]$ is a functional of
a {\it classical} field, having resulted from an
integration over microscopic degrees of freedom, we shall
propose that it plays the role of
the free energy associated with sigma condensates. 

From such a
premise, the bulk of our effort will be to establish direct connections
between the critical QGP system and observed patterns in hadron production. 
It is our aim to make specific predictions
relating to the extensive experimental programme on heavy ions now in
progress (SPS, RHIC, LHC), some of which have already been reported in
previous, brief, expositions \cite{ANT1}.
Our results will be summarized in an
assortment of Monte-Carlo simulations stemming from the advocated 
theoretical approach. Given that intermittency effects in hadron
production patterns are foremost in our minds, the significance of
the presented output should be directly relevant to the goal of 
identifying clear experimental signals for the existence of a critical point
in quark matter.

As can be surmised from the above remarks, our proposed scheme relies
centrally on the idea that the
observed patterns in hadron production have a component whose source is the
large scale dynamics induced by the underlying microscopic theory at the
critical point.
At the same time, it constitutes an effort which stays close to an equilibrium
frame of description\footnote{We shall not, in this paper, be preoccupied 
with precise dynamical accounts underlying the time development of the QGP
system from its initial formation to its critical stage. Our basic concern
is, simply, to justify our choices for the geometrical profile of the
QGP state as the critical point is approached.}, in juxtaposition with
alternative approaches which place primary emphasis on the non-equlibrium 
aspects of the problem, see, {\it e.g.} \cite{CDC}.

Clearly, we cannot altogether avoid the out-of-equilibrium dynamical 
component entering the transition from the (critical) QGP to the hadronic
state of the
considered physical problem. In our scheme this matter enters in 
connection with the critical slowing down expected to occur between the critical,
$T_c$, and freeze-out, $T_f$, temperatures - the latter marking the 
onset of the emergence of hadrons. Adopting a working assumption of the form 
$T_c\approx T_f$, owing to the fact that the specific heat diverges at the
critical point,
allows us to remain within the equilibrium frame of description.
On the other hand, 
the (proper) time interval between the aformentioned two temperatures is 
{\it not} necessarily negligible.
Clearly, dynamical matters related to critical slowing down effects
present an interest of their own. We shall not, 
in this paper enter the discussion of issues related
to dynamical scaling behavior.

Our paper is organized as follows. Section 2 is devoted to the
presentation of the field theoretical framework within which we shall
pursue the study of the critical system, assuming Bjorken's evolution
picture for the QGP. In Section 3  we develop a Statistical Mechanical
frame of
description for the critical field system which relies on cluster
formation. We proceed, in Section 4, to discuss the statistical content of
the global system, setting the stage for the development of a Monte-Carlo
generator for
critical events that will be taken up in Section 6.
In the interim,
Section 5, we shall display the basic battery of formulas pertaining to
the spherical evolution scenario for the QGP system. Section 6 addresses the 
quintessential issue regarding the generation of events with critical
fluctuations as well as their
identification in the profile of the produced
pions which can be traced to a critical point of second order.
In section 7 we present our numerical results by performing an
event-by-event analysis in a large set of critical events and showing how
our dedection algorithm of the critical fluctuations works in practice.
Finally, section 8 contains concluding remarks.

\section{Adjustment of the 3-D Ising-model to Bjorken's evolution 
scenario}

Our basic theoretical input is provided by the effective action of the
3-D Ising model $\Gamma [\sigma] $,  which, following the analysis  
of \cite{Steph},
adequately represents QCD in the vicinity of the second order critical point
$T=T_c,\mu=\mu_c$ (as the endpoint of a first-order phase transition line)
resulting by taking into account the finite mass for up and down
quarks. Thus, provided     
driving parameters of the model (e.g., temperature, coupling constant, etc.)  
are on or near their critical values, $\Gamma_c\left[\sigma\right]$ 
can be looked upon as the free energy for the critical 
system. 

According to our viewpoint the aformentioned free energy 
incorporates the sigma field
condensates, $\sigma \sim \left\langle \overline{q}q\right\rangle $,
in thermal equilibrium near the critical temperature $T \approx T_c$.
As already mentioned in the introductory section, for the most part of this
work we shall feel phenomenologically justified to keep our considerations 
on a strictly equilibrium level by taking into account that
the freeze-out temperature $T_f$,
at which one meets the asymptotic states of the produced hadronic system,
is very close to the critical temperature ($T_c\approx T_f$). This means 
that we shall view the second order phase as a direct interpolation between 
the condensates and the asymptotic hadronic states. 

Under conditions of strict local equilibrium the thermodynamics of the
condensates is described, in the ordered phase ($T \leq T_c$), by the
free energy which we express, at $T=T_c$, as follows
\begin{equation}
\Gamma_c[\sigma]=T_c^{-1} \int d^3 \vec{x}[\frac{1}{2}(\nabla \sigma)^2 +
g T_c^4 (T_c^{-1} \sigma)^{\delta +1}]
\label{eq:action}
\end{equation}
where $\delta$ is the isothermal critical exponent and $g$ is a
dimensionless coupling.
The field $\sigma$ has a dimension $\sigma \sim$ (length)$^{-1}$.
The parameters $(g,\delta)$ are universal and  specify the critical
equation of state:
$\frac{\delta \Gamma_c}{\delta \sigma} \sim g \sigma^\delta$.
For the Ising 3-D universality class we have $\delta \approx 5$, due
to the smallness of the anomalous dimension \cite{Wet}, and
$g \approx 1.5 - 2.5$ as obtained in \cite{Tsyp,Wet}.

Our next concern is to cast the effective action $\Gamma
_c\left[ {\bf \sigma }\right]$ in a manner that is explicitly adapted to 
Bjorken's
inside-outside cascade picture for the plasma expansion. For this purpose we
introduce rapidity and proper time coordinates $(\xi ,\tau )$ so that the
longitudinal space element, corresponding to a local observer in a comoving
system, becomes $dx_{\mid \mid }=\tau d\xi .$ With this choice and for the
purpose of describing the clusters formed at $T=T_c$, the ``longitudinal"
integration occurs along the critical hyperbola $\tau =\tau_c$.
We thereby obtain the following expression for the
effective action which furnishes an appropriate account for the system as    
the critical temperature is reached:

\begin{equation}
\Gamma _c\left[ {\bf \sigma }\right] =\frac{1}{C_A}
\int_\Delta d\xi \cosh \xi
\int_{S_{\perp }}d^2{\bf x}_{\perp }\left[ \frac {1}{2 \cosh \xi}\left(
\frac{\partial {\bf \sigma }}{\partial \xi }\right)^2
+\frac{\tau^2_c}{2}\left( \nabla_{\perp}{\bf \sigma }\right)^2
+g T_c^4C_A^2 (T_c^{-1} \sigma)^{\delta +1} \right]
\label{eq:action1}
\end{equation}
where $\Delta $ is the rapidity size, $S_{\perp }$ is the transverse area of
the system and $C_A=\frac{\tau_c}{\beta_c}$.

One notices that, through (\ref{eq:action1}), the system has been accomodated
on a
space-like hyperboloidal surface which is flat in the transverse directions 
$x,\,y$ and intersects the $t-z$ plane as a hyperbola ($z$ is the collision 
direction). The monitoring of the field system in our case 
calls for a collection of local inertial observers who must put their data 
together in order to construct its full description. This occurence is 
ideally suited to the Bialas-Peschanski scenario concerning the detection 
of intermittency patterns in the multiparticle hadronic production. Within
this picture it turns out that the total rapidity size $\Delta$ in
eq.(\ref{eq:action1})
should be replaced by a narrow strip $\Delta \xi$ in rapidity around a local
observer. Furthermore, we are free to choose $\Delta \xi \leq 1$ and make use
of the approximation $\cosh \xi \approx 1$ which is valid within 10 $\%$.

As already stated our central goal is, assuming that 
$T_c\approx T_f$, to formulate correspondences between field quantities 
associated with the critical system, on the one hand and hadronic
observables on the other. On the field 
theory side the relevant issue concerns the statistical account of the
macroscopic system comprised of condensates. One needs to define a
statistical weight factor appropriate for the equilibrium situation in
hand. A convenient way in terms of which one might proceed to accomplish this
task is to employ a coherent state analysis (see, e.g., \cite{CoSta}) which,
by definition, pertains 
to collective fields. Let us recall that coherent states are eigenstates
of field operators, being a superposition of Fock states of arbitrary
particle content. One writes (the {\it tilde} on top signifies an operator) 
\begin{equation}
\tilde{\sigma}(x)|\sigma>\,=\,\sigma(x)|\sigma>,
\end{equation}
where the $|\sigma>$ denote coherent (classical, in effect) states of the
field $\sigma$. Our basic assumption is that the density matrix associated
with the partition function is diagonal in the coherent-state representation.

In turn, the average multiplicity $<n>$ of sigmas is given by
\begin{equation}
<n>=\int{\cal D}[\sigma] \left[\int_Vd^2\vec{x}_\perp
d\xi\sigma^2(\vec{x}_\perp ,\xi)\right]e^{-\Gamma_c[\sigma]}/
\int{\cal D}[\sigma] e^{-\Gamma_c[\sigma]}
\label{eq:mulsig}
\end{equation}

The above formulas will be centrally relied upon in our subsequent
analysis. They, in fact, serve as a bridge between an original
microscopic description - which has been given a macroscopic, collective content
by integrating out a huge number of degrees of freedom -
and the statistical treatmeant of fluctuations in the resulting critical, 
collective system \cite{CoSta}.

\section{Statistical Mechanics of the critical system: cluster formation}

To the extent that we shall be restricting ourselves to a strictly
equilibrium mode of description, the strategy by which
we propose to study the critical system is through cluster formation. The
physical picture we attach to (critical) clusters is that they represent 
distributions of primordial hadrons (massless sigmas) whose
growth is driven by the transition from the QGP
to the hadronic state of matter.
From a formal, mathematical, viewpoint such clusters serve as entities
which interpolate between critical exponents characterizing the plasma system
on the one hand and fractal dimensions associated with the hadronic (pion)
emission patterns on the other.

The 3-D configurations representing the clusters will be assembled as follows.
First, their one-dimensional (longitudinal) profile along the rapidity
axis will be considered, given that this is
where the bulk of the dynamical effects is encountered. Next, we shall
proceed to investigate their transverse fractal properties by freezing the 
rapidity variable. The cartesian synthesis of a given cluster,
provides a well defined picture of a cluster's constitution. The latter,
certainly adheres to 
the cylindrical (with respect to the rapidity axis) and Lorentz
(save at the edges) symmetries which characterize the overall system.

The projection onto the rapidity direction is furnished by
the 1-D effective action functional:
\begin{equation}
\Gamma_c^{(1)} [ \hat{\sigma} ]=\frac{\pi R_{\perp}^2}
{\beta_c \tau}
\int_{\Delta \xi} d\xi \left[ \frac{1}{2} \left(\frac{\partial
\hat{\sigma}}
{\partial \xi}\right)^2
+g C_A^2
\left(\hat{\sigma}^2\right)^{\frac{\delta +1}{2}} \right]
\label{eq:rapproj}
\end{equation}
where $\Delta \xi$ is the rapidity strip around the local observer at $\xi=0$
and $R_{\perp}$ the transverse radius of
the system at $T=T_c$, while $\hat{\sigma}$ denotes a dimensionless
$\sigma$-field ($\hat{\sigma}=T_c^{-1} \sigma$).

For notational ease we make the identifications
\begin{equation}
g^{(1)}_1=\frac{\pi R_{\perp}^2}{\beta_c \tau_c}~~~~;~~~~g^{(1)}_2=
g C_A^2
\label{eq:1dcoupl}
\end{equation}
which recast Eq.(\ref{eq:rapproj}) into the following generic form
\begin{equation}
\Gamma_c^{(1)} [ \hat{\sigma} ]=g_1^{(1)}
\int_{\Delta \xi} d\xi \left[ \frac{1}{2} \left(\frac{\partial
\hat{\sigma}}
{\partial \xi}\right)^2
+g_2^{(1)} \left(\hat{\sigma}^2\right)^{\frac{\delta +1}{2}} \right]
\label{eq:genact}
\end{equation}

The local system is considered as {\it open}, i.e. it communicates with
the total system, so that no boundary conditions are imposed on it.
In turn, this implies that we are free to consider analytic continuation of
(\ref{eq:genact}), with respect to $\Delta \xi$,
which will furnish
scaling properties of extensive thermodynamical quantities registered by
the local (sub)system as
its size is being varied. We emphasize that our specification $\xi=0$ has
been made strictly for notational ease; any local oberver, located at
$\xi=\xi_i$ is equally acceptable, due to Lorentz invariance, provided
that the said observer is not too close to the boundaries of the total
system.

For heavy nuclei, $\frac{\pi R^2_{\perp}}{\beta_c \tau_c}\gg 1$,
the evaluation
of the partition function of the subsystem
\begin{equation}
Z_c\,=\int {\cal D}[\hat{\sigma}] e^{-\Gamma_c[\hat{\sigma}]}
\end{equation}
is dominated by saddle point configurations.
In what follows we drop, for simplicity of notation, the
hat used to indicate dimensionless quantities.

There emerge instanton-like solutions
of the classical equations of motion \cite{PRL}
\begin{equation}
\ddot{\sigma}-(\delta +1) g_2^{(1)} \sigma^{\delta}=0
\label{eq:eqmot1d}
\end{equation}
which are classified according to total energy
\begin{equation}
E=\frac{1}{2}\dot{\sigma}^2-g^{(1)}_2|\sigma|^{\delta+1}
\end{equation}
and size $\xi_0$.

Now, for  a given solution it follows, in a straight forward manner, that
a suppression term
$e^{-E\Delta\xi}$ factorizes in the expression for the partition function,
hence the dominant contributions correspond to configurations with
vanishing energy. Under these circumstances one obtains the following
analytic expression for the instanton-like solution
\begin{equation}
\sigma(\xi)\,=\,\left[\frac{\sqrt{2}}{(\delta-1)\sqrt{g_2^{(1)}}}
\right]^{\frac{2}{\delta-1}}|\xi-\xi_0|^{-\frac{2}{\delta-1}} 
\label{eq:solut1d}
\end{equation}

As a function of the varying size $\Delta\xi$ the effective action takes
the form 
\begin{equation}
\Gamma_c(\Delta\xi;\xi_0)\,=\,\frac{2 g \pi R^2_\perp}{\beta_c \tau_c}
C_A^2\int_{\Delta\xi}[\sigma(|\xi-\xi_0|]^{\delta +1}d\xi 
\label{eq:eff1d}
\end{equation}

In order to suppress large contributions to the free energy, hence
practically zero contributions to the partition function, we impose the
restriction that the ``instanton" size be much larger than the size of the
local system, i.e. $\Delta\xi \ll\xi_0$. As a result, we obtain
approximately constant solutions in the domain $\Delta\xi$, of the form
\begin{equation}
\sigma \approx \left[\frac{\sqrt{2}}{\xi_0 \sqrt{g_2^{(1)}(\delta-1)}}
\right]^{\frac{2}{\delta-1}} 
\label{eq:sigcon}
\end{equation}

The computation of the partition function is now reduced to an integration
over the ``instanton" size. We let physical insight guide the relevant
calculation by thinking as follows. Consider an extensive variable $M$
associated with the field configuration of the critical system in the
region monitored by our local observer, including its analytic
continuation. Specifically, we choose the quantity
$M=\int_0^{\Delta\xi}[\sigma(x)]^2dx$
which furnishes, according to eq.(\ref{eq:mulsig}),
the multiplicity $<n(\Delta\xi)>=<\int_0^{\Delta\xi}
[\sigma(x)]^2dx>$ of sigmas
within the domain of extent $\Delta\xi$. Moreover, we introduce the
concept of a {\it cluster} of radius $\Delta\xi$ as an object 
with geometrical properties 
built up through the statistical average over configurations corresponding
to values of $M$ greater than a minimum value $\mu$ (for the case at hand
$\mu=1$).
Referring to (\ref{eq:mulsig}) and (\ref{eq:eff1d})
this average is determined as follows
\begin {eqnarray}
<\int_0^{\Delta\xi}[\sigma(x)]^2dx>&=&\frac{A^2 \frac{\delta-1}{\delta-3}}{Z}
\int_{\Delta \xi}^{(A^2 \Delta \xi/ \mu)^{\frac{\delta-1}{4}}}d\xi_0\,
\xi_0^{-\frac{\delta+1}{\delta-1}}\left[\xi_0^{\frac{\delta-3}{\delta-1}}- 
(\xi_0-\Delta\xi)^
{\frac{\delta-3}{\delta-1}}\right]\nonumber\\& &\quad\times {\rm
exp}\left\{-G_1 \frac{\delta-1}{\delta+3}
\left[(\xi_0-\Delta\xi)^
{-\frac{\delta+3}{\delta-1}}-
\xi_0^{-\frac{\delta+3}{\delta-1}}\right]\right\}
\label{eq:magmean}
\end{eqnarray}
where $G_1\equiv 2g^{(1)}_1g^{(1)}_2A^{\delta+1}$
and $A=[g^{(1)}_2 /2(\delta -1)^2]^{-\frac{1}{\delta -1}}$.

It can now be explicitly shown \cite{PRL} that,
for $G_1\gg 1$, three characteristic
regions can be clearly identified with respect to the behaviour of the
integral on the right hand side of (\ref{eq:magmean}):

\begin{eqnarray}
\Delta\xi \ll \Delta_d~~~~~~&;&~~~~~~<n(\Delta\xi)> \sim const. \nonumber
\\
\Delta_d \ll \Delta\xi \ll \Delta_u~~~~~~&;&~~~~~~
<n(\Delta\xi )> \sim
\left(\Delta\xi\right)^{\frac{\delta-1}{\delta+1}}\nonumber\\
\Delta\xi \gg \Delta_u~~~~~~&;&~~~~~~
<n(\Delta\xi )> \sim \left( \Delta\xi \right)^{\frac{\delta-5}{\delta-1}}
\label{eq:plaws1}
\end{eqnarray}
where
\begin{equation}
 \Delta_d\equiv A^{-2 \frac{\delta+1}{\delta-1}}G_1^{\frac{2}{\delta-1}}
\mu^{\frac{\delta+1}{\delta-1}}~~~;~~~\Delta_u\equiv
G_1^{\frac{\delta-1}{\delta+3}}
\label{eq:rapsiz}
\end{equation}
serve as the lower ($\Delta_d$) and the upper radius ($\Delta_u$)
of a cluster centered around $\xi=0$.

There follows a fractal structure for the critical clusters with mass
dimension
\begin{equation}
d_F^{(1)}=\frac{\delta-1}{\delta+1}
\end{equation}

Turning our attention to the two-dimensional, transverse profile of the
critical system we encounter the effective action functional
\begin{equation}
\Gamma_c^{(2)} [ \sigma ]=C_A \Delta 
\int_\perp  d^2 x_\perp \left[ \frac{1}{2} (\nabla_\perp
\sigma_{\perp})^2
+ g 
\left(\sigma^2 \right)^{\frac{\delta+1}{2}} \right]
\label{eq:2dproj}
\end{equation}
where $\perp$, as subscript to the integral, serves to denote transverse
size and $\Delta$ furnishes the total rapidity. 

The form of the above expression is a two-dimensional analogue of
(\ref{eq:genact}) with corresponding parameters
\begin{equation}
g^{(2)}_1=C_A \Delta~~~~;~~~~g^{(2)}_2=g 
\label{eq:2dcoupl}
\end{equation}

The counterpart of eq.(\ref{eq:plaws1}) now reads
\begin{eqnarray}
R \ll R_d~~~~~~&;&~~~~~~<n_{\perp}(R)> \sim const.  \nonumber \\
R_d \ll R \ll R_u~~~~~~&;&~~~~~~<n_{\perp}(R)> \sim R^{2 \frac{\delta-1}
{\delta +1}}  \nonumber \\
R \gg R_u~~~~~~&;&~~~~~<n_{\perp}(R)> \sim R^{2-\frac{4}{\delta-1}}
\label{eq:plaws2}
\end{eqnarray}
where
\begin{equation}
R_d=G_2^{\frac{1}{2 \delta}}A_2^{\frac{\delta+1}
{2 \delta}}~~~;~~~R_u=\beta_c G_2^{\frac{\delta-1}{4}}
\end{equation}
with $G_2=2 \pi g_1^{(2)} g_2^{(2)} A_2^{\delta +1}
\left(\frac{\delta +3}{4}\right)$
and $A_2=((g_2^{(2)}/4)(\delta -1)^2 (\delta +1))^{-\frac{1}{\delta -1}}$.
In analogy with
eqs.(\ref{eq:plaws1}) we conclude from eqs.(\ref{eq:plaws2}) the formation of 
fractal clusters in transverse space with mass dimension $d_F^{(2)}=
\frac{2 (\delta - 1)}{\delta + 1}$ in the range $R_d \ll R \ll R_u$.
The 3-D cluster is constructed
following our original considerations according to which this object
is roughly confined within a cylinder with radius
$R_u$ in transverse space and
size $2 \Delta_u$ in rapidity. This realization has
the cartesian product form (rapidity $\bigotimes$ transverse space)
and it is consistent with longitudinal expansion of the original system.

Referring to eq.(\ref{eq:mulsig}),
with $\Gamma_c[\sigma]$ given by (\ref{eq:rapproj}),
we estimate the average
multiplicity for the 3-dimensional cluster to be \cite{PRLN}:
\begin{equation}
<n>_{cl}=\frac{\Gamma(\frac{3}{\delta +1})}{\Gamma(\frac{1}{\delta +1})} 
\left(\frac{V}{V_o}\right)^{\frac{\delta -1}{\delta +1}}
\label{eq:clusmul}
\end{equation} 
with $V=2 \pi R_u^2 \Delta_u$ and $V_o=\beta_c^2
\sqrt{{2 g C_A}}$.

\section{Multi-cluster description of the global system}

Assuming that the entire critical system is a cylinder with rapidity
size $\Delta$ and transverse radius $R_{\perp}$ we can calculate the
mean number of `cylindrical' clusters with rapidity size
$2 \Delta_u$
and transverse radius $R_u$ contained in the global system.
Denoting this number
by $N$ we determine
$N=N_{\parallel} \times N_{\perp}$, where $N_{\parallel}$ denotes the
number of
clusters in rapidity and $N_{\perp}$ 
in transverse space, respectively. Using the fact that 
$\kappa=\frac{\delta +1}{2} \approx 3$
for the Ising 3-D universality class \cite{Tsyp,Wet} we estimate, from
the previous analysis, these numbers as:
\begin{eqnarray}
N_{\parallel}&=&\frac{\Delta}{2 \Delta_u} = \frac{2 \Delta \tau}
{\sqrt{\pi} R_{\perp}} \left(\frac{g}{2}\right)^{1/4} ,\nonumber \\
N_{\perp}&=&\left(\frac{R_{\perp}}{R_u}\right)^2 =
\left(\frac{12 \sqrt{6 g} R_{\perp}}{\pi \Delta \tau}\right)^2  
\label{eq:numclus}
\end{eqnarray}

It follows that the 3-D critical system appears as an almost cylindrically
symmetric\footnote{We say `almost'
because the individual clusters are not supposed to behave as rigid
structures;
they can deform, due to the presence of neighbouring clusters, both in
transverse and in rapidity directions.}
arrangement of, in the mean, $N$ clusters. 
Each cluster
(of size $2 \Delta_u$ in rapidity and $R_u$ transverse radius) has a
fractal
mass dimension $d_F$ decomposed as the cartesian product
$d_F=d^{(1)}_F + d^{(2)}_F$ of two fractals in the corresponding 
subspaces. We also observe that the size of the clusters in rapidity is
relatively small, $\Delta_u = \frac{\sqrt{\pi}}{4}\left(\frac{R_{\perp}}
{\tau}\right)$, justifying a posteriori the approximation $\cosh \xi \approx
1$ made in eq.(\ref{eq:action1}).

The complete analysis calls for the power
law underlying the behavior of the
density-density correlation function $\rho(\xi,0)=<\rho(\xi)\rho(0)>$ 
in the scaling region. Differentiating the mean multiplicity, given by
eq.(22), with respect to rapidity we are led \cite{PRLN} to
\begin{equation}
\rho(\xi,0) =
(\frac{4}{27 \pi g C_A})^{1/3} \beta_c^{-2} \frac{\Gamma(1/2)}{\Gamma(1/6)}
\vert \xi \vert^{-1/3} \vert\frac{\vec{x}_{\perp}}{\beta_c} \vert^{-2/3}  
\label{eq:dedeco}
\end{equation}
As already mentioned, the fractal dimension emerging from the power-laws
in rapidity space is
given by $d_F^{(1)}=\frac{\kappa-1}{\kappa}$ and is valid in the scaling
region $ \delta_o(\equiv\Delta_d)\ll |\xi|\leq \Delta_u$.

In the presence of $N_\parallel$ non-overlapping sources, located along
the rapidity direction at random points $\xi_1,...,\xi_{N_\parallel}$, the
corresponding partition function reads:
\begin{equation}
Z_{N_\parallel}=\sum_{\xi_1,...,\xi_{N_\parallel}}
\prod^{N_\parallel -1}_{i=1}
\left[-\delta_o^{\kappa-1}\left(\frac{\pi R_\perp^2}
{\beta_c^2}\right)^\kappa
\int_{\Delta\xi_i}d\xi(<\sigma^2>)^\kappa\right],
\end{equation}
which leads \cite{Anton} to the following distribution for the given multiplicity
$N_\parallel$:
\begin{equation}
P(\xi_1,...,\xi_{N_\parallel})=\left(\frac{\Delta+2\xi_1}{\delta_o}
\right)^{-\frac{1}{2\gamma}}\left(\frac{\Delta-2\xi_{N_\parallel}}{\delta_o}
\right)^{-\frac{1}{2\gamma}}\prod_{i=2}^{N_\parallel}\left(\frac{\xi_i-
\xi_{i-1}}{\delta_o}\right)^{-\frac{1}{\gamma}}
\label{eq:glosy}
\end{equation}
where $\gamma\equiv\left[\frac{9}{2}\Gamma\left(\frac{2}{3}\right)\right]^3$,
with the assumed ordering being $-\frac{\Delta}{2}\leq\xi_1
\leq...\leq\xi_{N_\parallel}\leq\frac{\Delta}{2}$. One readily notes that the 
largeness of $\gamma$ allows us to regard the chain of clusters along the
rapidity direction as a collection of practically non-interacting objects.
This occurence will greatly facilitate, later on, our numerical simulations.

Unlike the rapidity case, fluctuations associated with
transverse
space are not directly observable in heavy ion experiments. What
actually {\it is} observed is the pion distribution in the transverse
momentum
space. The need, therefore, arises to transform the geometrical picture
valid for the transverse space to the corresponding picture in transverse
momentum space.
The two spaces are connected at the level of the density-density
correlation through a Fourier transform:
\begin{equation}
\tilde{\rho}(\vec{p}_T) \sim \int d^2 x_{\perp}
e^{i \vec{p}_T \cdot \vec{x}_{\perp}}~\rho(\vec{x}_{\perp},0) 
\label{eq:foutr}
\end{equation}
where $\rho(\vec{x}_{\perp},0)$ is the density-density correlation
for the sigmas in the transverse space projection of a critical
cluster with center at $\vec{x}'_{\perp}=0$ and
$\tilde{\rho}(\vec{p}_T)$ the corresponding density-density correlation
in the transverse momentum space. The latter
is, by definition, related to the gradient
of the mean multiplicity $<n_{\perp}(R)>$, see eqs.(\ref{eq:plaws2}),
with respect to $R$. 

Setting $R=\vert \vec{x}_{\perp} \vert$ we obtain
\begin{equation}
\rho(\vec{R},0) \sim \frac{1}{R} \frac{\partial <n(R)>}{\partial R}
\sim R^{-\frac{2}{\kappa}}
\end{equation}
Performing the Fourier integration, according to (\ref{eq:foutr}),
we arrive at the following power-law
for the density-density correlation in the transverse momentum space:
\begin{equation}
\tilde{\rho}(\vec{p}_T) \sim \vert \vec{p}_T \vert^{\frac{2 (\kappa-1)}
{\kappa}}
\end{equation}  
In analogy to the space description of the 3-D system, we now consider a
collection of cylindrical clusters in the transverse momentum
$\bigotimes$ rapidity space.
The number of clusters in the transverse momentum projection is
$N_{\perp}$ given by eqs.(\ref{eq:numclus}).
Finally, the multiplicity
of hadrons within one cluster is given by eq.(\ref{eq:clusmul}).

In order to
complete the description of the critical system in the transverse
momentum we need an additional phenomenological input.
We adopt the following assumptions: 

\begin{itemize}

\item{The centers of the clusters in the transverse momentum space
$p_{T,i}$
are random variables distributed according to an exponential law:
\begin{equation}
f(p_T)=\frac{2}{\pi <p_T>^2} e^{-\frac{2 p_T}{<p_T>}}
\label{eq:explaw}
\end{equation}}

\item{The net transverse momentum flow through the clusters is suppressed:
\begin{equation}
\sum_{i=1}^{N_{\perp}} \vec{p}_{T,i} = 0
\label{eq:momcons}
\end{equation}}
\end{itemize}

\noindent where $<p_T>$ is the mean transverse momentum
which can be approximated to:
$<p_T> \approx 2 T_c =280~MeV$. The size $\vert \Delta \vec{p}_{\perp} \vert$
of each cluster in the transverse
momentum space is determined by the minimum length scale $R_d$ in the
transverse configuration space and is given by the relation \cite{PRLN}:
\begin{equation}                                             
\vert \Delta \vec{p}_{\perp} \vert = \frac{T_c}{2}
(\frac{R_{\perp}}{\tau})^{1/2} (\frac{g}{2})^{-3/8}
(\frac{\pi^3}{C_A})^{1/4}
\label{eq:trmrad}
\end{equation}

Having exhibited the portrait of the critical system according to the 
longitudinal evolution scheme, we can now turn our attention 
to the spherical mode of expansion. This we shall do in the following 
section through a display of a minimum of formalism.

\section{Spherical evolution}

In the present section we shall consider a different evolution scenario
according to which the geometry of the $A+A$ collision is spherical. More
specifically, we consider the situation where, following the collision, a
fireball of spherical size is formed which expands radially outwards.
As a result, the sigmas are now distributed inside a spherical cluster.

The starting point of our analysis is, once again, the 3-D effective
action $(1)$.
Following a similar procedure with the one applied for the 
cylindrical geometry we find that
the average multiplicity $<N_{\sigma}>$ is given by
\begin{equation}
<N_{\sigma}>=Z^{-1} \int {\cal{D}}[\sigma] e^{- \Gamma_c[\sigma]}
(\beta_c^{-1} \int d^3 r \sigma^2(r))
\label{eq:defmult}
\end{equation}
where the insertion of the factor $\beta_c^{-1}$ has been made on
dimensional
grounds since $dim[\sigma^2] = (length)^{-2}$ and $<N_{\sigma}>$ must be
dimensionless.

The partition function at $T=T_c$, with the scalar particles ($\sigma$) in
a spherical volume $V (\leq (4/3) \pi R_u^3)$, now has the form
\begin{equation}
Z(N_{\sigma}, V, T_c)=N_{\sigma}^{-1/2}
\exp(-(\frac{V}{V_o})^{\kappa -1} N_{\sigma}^{\kappa})
\label{eq:part}
\end{equation}
where $V_o= (2g)^{\frac{1}{\kappa -1}} \beta_c^3$. 

The upper limit $R_u$ of the cluster can be computed directly from
\cite{PRL}. For the $D=3$ case the anomalous dimension $\eta$ of the
sigma-field is involved in the resulting expression.
We obtain $R_u=G_3 ^{-\frac{1}{q}}$ with
$q=3- \frac{(2+3 \eta)\kappa}{2}$ and $G_3=\sqrt{\frac{2}{g_2^{(3)}}}
\frac{\pi}{3} g_1^{(3)}$ \cite{PRL}. 
For small $\eta$ ($\eta \ll 1$) we find that $R_u \to \infty$.
Accordingly, we can
consider the whole system as being comprised of only one cluster with
radius $R_u$.
The average multiplicity $<N_{\sigma}>$ in this cluster, as a
function of the volume $V$, follows from (\ref{eq:part}) and is given by
\begin{equation}
<N_{\sigma}>=
\frac{\Gamma(\frac{3}{2 \kappa})}{\Gamma(\frac{1}{2\kappa})}
\left(\frac{V}{V_o}\right)^{\frac{\kappa-1}{\kappa}}
\label{eq:mults}
\end{equation}
The above equation implies that there is a minimal volume scale,
$V=V_o$, where self-similarity breaks down. 

On general grounds, the fractal dimension in $R^3$-space is given by 
$d_F=3
\frac{\kappa-1}{\kappa}$ if we neglect $\eta$.
Taking $\eta$ into account ($\approx 0.034$ \cite{Wet})
we obtain
$d_F=1.98$. Thus if we were to project this fractal onto rapidity and
transverse space, respectively, we would not
find corresponding fractal configurations since the dimensions of the
spaces upon which we project are smaller or nearly equal to $d_F$ \cite{FRAC}. 

To complete our considerations in configuration space we give 
the expression for the
density-density correlation function which obeys 
the following characteristic power-law
\begin{equation}
<\rho(r) \rho(0)>= \frac{\Gamma(\frac{3}{2 \kappa})}
{\Gamma(\frac{1}{2 \kappa})}(\frac{8 \pi g}{3})^{-\frac{1}{\kappa}}
\beta_c^{\frac{3- 2\kappa}{\kappa}} r^{-\frac{3}{\kappa}}
\label{eq:dens}
\end{equation}

Turning our attention to momentum space, the pattern is again a spherical
cluster with center at $\vec{p}=0$. The corresponding radius is determined by
the lower limit $R_d$ of the cluster size in configuration space, {\it i.e.}
$\vert \Delta \vec{p} \vert \sim R_d^{-1}$. On the other hand, $R_d$ is
determined by the minimal volume $V_o=(4/3)\pi R_d^3$. It follows
that:
\begin{equation}
\vert \Delta \vec{p} \vert \approx
\left(\frac{2^{\frac{2 \kappa-3}{\kappa-1}} \pi}
{3 g^{\frac{1}{\kappa-1}}}\right)^{1/3}
\beta_c^{-1}
\label{eq:3dmsiz}
\end{equation}
 and for $g \approx 2$, we estimate
$\vert \Delta \vec{p} \vert \approx 150~MeV$. The fractal dimension in
momentum space is $\tilde{d}_F=3-d_F=3(1- \frac{\kappa-1}{\kappa})$ and,
the momentum distribution of sigmas inside
the cluster is determined through the density-density correlation:
\begin{equation}
<\tilde{\rho}(\vert \vec{p} \vert) \tilde{\rho}(0)>
\sim \vert \vec{p} \vert^{- 3 \frac{\kappa-1}{\kappa}}
\label{eq:plaws3}
\end{equation}
which results
from the Fourier transform of eq.(\ref{eq:dens}).

This concludes our statistical mechanical
account for the critical fluctuations in terms of cluster formation, for 
each of the two alternative evolution schemes. We shall proceed, in the
next section, to discuss corresponding numerical studies,
pertaining to a critical component for pion
production in A+A collisions through simulations of a Monte-Carlo type.
Our expectation is that  the `theoretical generation' of events whose
origin lies
exclusively in a critical state reached by the QGP system, will aid
the search for respective signals in produced multiparticle pion states.

\section{Monte-Carlo simulations}

Following the considerations in the previous sections it becomes obvious
that the formation of critical clusters is expected to induce direct
phenomenological consequences, in the multiplicity distribution of the
produced sigmas, in the final states of a $A+A$ collision. In the first
part of the present section we shall describe the
development of a Monte-Carlo programme for the purpose of simulating the
production of ``critical pions" in such collision processes,
coming from the decay of light sigmas. The reader who is not interested
for the technical details of our approach can skip the first subsection and
go directly to the next one where we discuss a proposed algorithm for 
detecting the critical fluctuations in experimental data.

\subsection{CMC Generator for critical events}
In the
context of the QCD phase diagram in the temperature/chemical potential
plane, a two step procedure is required for the study of the critical pion
spectrum formation.
First, we need to produce the
clusters corresponding to the $\sigma$-condensates. According to our
theoretical discussion, the geometry of the
$\sigma$-clusters is determined by the evolution pattern of the
centrally formed fluid of quarks and gluons. In the case of
longitudinal evolution we have a cylindrical arrangement of several clusters
in the transverse momentum $\bigotimes$ rapidity space (cf. sections 3 and 
4) while in a spherically symmetric evolution a single cluster in momentum
space is formed (cf. section 5).

Second, we let the sigmas in the condensates decay
into pions with a branching ratio 1:2 for neutral to charged pion
production. The crucial parameter for determining the spectrum of
the produced pions is provided by the mass of the decaying sigmas. As the
sigma mass is not a constant but, {\it in fact}, evolves during the
freeze-out process, we treat it as a varying parameter in our numerical
considerations. 

We recall that the number
of clusters in the case of cylindrical symmetry is determined by the
size of the critical region while the multiplicity of $\sigma$'s within
each cluster depends on the couplings occuring in the free
energy of the system at the critical point.  Once the sigma mass is
assigned a specific value, we use
two-body phase space kinematics in order to determine the transverse
momenta and the rapidities of the pions.
For the spherically symmetric
case, on the other hand, a single critical cluster is formed around the CM
of the system ($\vec{p}=0$) while
the multiplicity within the cluster depends, once again, on the
couplings of the ``critical'' free energy. In what follows we shall
conduct numerical investigations, using Monte-Carlo methodology, for both
the cylindrical and the spherical configurations of the critical clusters.
It could very well be that the geometry of the evolving critical system
is neither perfectly cylindrical nor spherical but consistent with a
picture interpolating between these two extreme regimes.

Let us consider the implementation of the first step (simulation of the
sigma-condensates) for a longitudinally
expanding critical system (cylindrical geometry). The input parameters
determining the sigma-distribution are: the size $R_{\perp}$ of the system
in the
transverse direction, the total rapidity interval $\Delta$, the proper time
scale $\tau$,
the critical temperature $T_c$, the coupling $g$ in the 3-D Ising
effective
free energy and the isothermal critical exponent $\delta$. Once we
fix the values of the input parameters
we can calculate the size of a
cylindrical cluster of sigmas, meaning its transverse momentum
radius $\vert \Delta \vec{p}_{\perp} \vert$, given by eq.(\ref{eq:trmrad}),
and its size in rapidity $2 \Delta_u$ (see eq.(\ref{eq:rapsiz})).
The number of sigma clusters in the
transverse (momentum or real) space projection of the critical system, 
as well as the corresponding number of clusters in
rapidity space are then given through eqs.(\ref{eq:numclus}).
Finally, the multiplicity of sigmas within each
cluster can be obtained using eq.(\ref{eq:clusmul}).

Next, we construct the phase space projections of the sigmas in
the critical event. The determination of the rapidity
space profile proceeds as follows:
We produce a configuration of $N_{\parallel}$ centers for these clusters
treating them as random variables, uniformly distributed in
$[0,\Delta]$ according to eq.(\ref{eq:glosy}). Then we generate around
each center the sigmas, distributed
with a density-density correlation following the power-law (\ref{eq:plaws1}).

This stage of the algorithm requires the construction of a random fractal
with a given fractal dimension $d_F$. In order to perform this construction
we have to use test-functions of the form $P(x) \sim x^{-1-d_F}$ as a
generalization of the L\'evy distributions for arbitrary embedding
dimension \cite{Levyf}. We produce randomly the rapidities corresponding to
one cluster using the appropriate test-function. In general,
the resulting set
does not have the correct center and size. Due to self-similarity the
cluster size is obtained through a simple scaling while, with a suitable
translation, the cluster is placed around the corresponding center.
Coalescence of neighbouring clusters
{\em is}
allowed, so
the actual multiplicity within each cluster depends on the position
of
the centers of its neighbours. Therefore, the cluster arrangement
in the rapidity space
determines the total multiplicity $N_{\sigma}$ of sigmas in the
event.

Turning our attention to the transverse momentum space we arrange
the centers of the $N_{\perp}$ clusters so that they are distributed
according to the exponential law (\ref{eq:explaw}),
under the constraint that the total transverse momentum of the
centers vanishes. To each such cluster correspond
$\frac{N_{\sigma}}{N_{\perp}}$ sigma-particles distributed according
to the power-law (\ref{eq:plaws2}). The construction algorithm for the
transverse momentum clusters is analogous to the algorithm used for the
construction in rapidity space. The test-function has to be adjusted to
produce the random fractal with the correct dimension.

We, therefore, end up with two sets of
phase space variables.
A set $S_T$ with $N_{\sigma}$ transverse momentum variables
$ \{ (p_{{T,x}_1},p_{{T,y}_1}),~(p_{{T,x}_2},p_{{T,y}_2}),...,~
(p_{{T,x}_{N_{\sigma}}},p_{{T,y}_{N_{\sigma}}}) \}$ and a set $S_{\xi}$
with $N_{\sigma}$ rapidity variables $\{
\xi_1,~\xi_2,...,~\xi_{N_{\sigma}}
\}$. The sigma-content of a ``critical'' event, respecting the
cartesian product form of the underlying cylindrical geometry, is then
realized as any possible one-to-one pairing of the elements of $S_T$ with
the elements of $S_{\xi}$. 

For a spherically symmetric expansion the size of the critical cluster
is represented by a single parameter $R$ corresponding to
the radius of the (growing) system in the 3-D space. This makes the
case in hand much easier to handle. First we specify
the size $\vert \Delta \vec{p} \vert$ of the (single) critical sigma cluster
in momentum space, cf. eq.(\ref{eq:3dmsiz}) and then using
eq.(\ref{eq:mults}) we determine the total sigma multiplicity
$N_{\sigma}$ within this cluster. The generation of the momenta of the
sigma-particles
is then based on the observation that for the 3-D system the 
density-density correlation function in momentum space coincides with the
step
distribution of a L\'evy flight \cite{Levyf} leading to a fractal set with
the same
fractal dimension ($\tilde{d}_F^{(3)} \approx 1$).
Thus the algorithm used for the generation of the $\sigma$-momenta is given
through the following equations
\begin{eqnarray}
p_{x,i}&=& p_{x,i-1} + \vert \Delta\vec{p} \vert r_1 \sqrt{1-r_2^2}
\cos(2 \pi r_3) \nonumber \\
p_{y,i}&=& p_{y,i-1} + \vert \Delta\vec{p} \vert r_1 \sqrt{1-r_2^2}
\sin(2 \pi r_3) \nonumber \\
p_{z,i}&=& p_{z,i-1} + \vert \Delta\vec{p} \vert r_1 r_2
\label{eq:ransph}
\end{eqnarray}
where $r_1$ is distributed according to the density-density correlation
function
(\ref{eq:plaws3}), $r_3$ is random variable uniformly distributed in $[0,1]$
while $r_2$ is uniformly distributed in $[-1,1]$. A final scaling and
translation, as in the case of the cylindrical clusters, is needed in
order to embed the cluster in a region with radius $\vert \Delta \vec{p}
\vert$ around $\vec{p}=0$ in momentum space.
 
In order to establish contact
with the observables in an experiment with colliding relativistic heavy
ions
we must take into account the decay of sigmas into pions. As already
discussed
this constitutes a straight forward step for a given value of the mass
of the decaying
sigmas. If this mass is far from the pion
production threshold $2 m_{\pi}$ the kinematics of the pions is strongly
influenced by the sigma-mass and the fluctuation pattern of the pions is
disordered relative to the spectrum of the sigmas.
On the other hand, the sigma mass is generated dynamically during the
freeze-out phase and, if the decay rate is much larger than the expansion 
rate, there is a good
possibility that the sigmas decay immediately after their effective mass
has reached the pion production threshold \cite{PRLN}. If this is the case
then the fluctuation patterns describing the sigma momenta are transfered,
almost unchanged, to the momenta of the produced pions. Here however, we will
follow a more general mechanism for the decay of the sigmas allowing us
also to
describe
a more conservative scenario when the sigmas decay well above the two
pion threshold. Specifically,
in order to simulate the variation of the sigma mass during the freeze-out
process, we introduce a probability distribution $P(M_{\sigma})$ determining
the number of sigmas with mass $M_{\sigma}$ which decay into pions. 
We assume that $P(M_{\sigma})$ is well described by a Gaussian of mean value
$M_{\sigma}$ and variance $\delta m_{\sigma}$.

The generation of the critical events is therefore completed in the last
stage
of our Monte-Carlo code (CMC) when, treating the sigma mass as a random
variable with propability density $P(M_{\sigma})$, we produce the final
charged pions through the decay of the critical sigmas. The quantum
mechanical amplitude describing the probability of the sigma-decay into a
$\pi^+$-$\pi^-$ pair is supposed to be independent of the sigma-momentum.

\subsection{Reconstruction of the critical $\sigma$-sector}

Using the CMC algorithm we produce a large number of critical events and we
study their phenomenology. Our main interest is to
reveal the critical fluctuations from the momenta of the final observed
charged pions. For this purpose we perform an event-by-event intermittency
analysis in the factorial moments of the negative pion momenta looking for
a region with self-similar fluctuations. If the mass parameter $M_{\sigma}$
is less than $1~MeV$ above the two pion threshold the critical fluctuations
can be directly observed in the momenta of the negative pions.
Increasing $M_{\sigma}$
the effect gets more and more suppressed leading, relatively soon
($M_{\sigma} \approx 10~MeV~+~2 m_{\pi}$), to an almost dissapearance of the
critical fluctuations (see Fig.~5, open up triangles). This behaviour
remains even if the variance $\delta m_{\sigma}$ vanishes.
If we increase $\delta m_{\sigma}$ for a given value of
$M_{\sigma}$ we find, in general, a suppression of the
critical region with rate depending on the value of $M_{\sigma}$. If
$M_{\sigma} \leq 1~MeV~+~2 m_{\pi}$ the rate of suppression is very small and the
critical fluctuations, for realistic values of $\delta m_{\sigma}$,
practically persist. In order to get back the critical behaviour one has,
therefore, to
reconstruct the critical sigmas based on the momenta of the observed charged
pions. We propose the following reconstruction algorithm:
\begin{itemize}
\item{For a given event, using as input the momenta of the charged pions,
form the invariant mass $m^2_{\pi^+ \pi^-}=(p_+ + p_-)^2$ for all possible
pairs of $\pi^+$-$\pi^-$ ($p_+$ ($p_-$) is the 4-momentum of $\pi^+$
($\pi^-$)).}
\item{Determine the probability distribution of the quantity
$m^2_{\pi^+ \pi^-}$. Look for a peak in the distribution. Form a new set of
charged pions picking up the
$\pi^+$-$\pi^-$ pairs within a region of a few $MeV$ ($\approx 50\% \delta
m_{\sigma}$) around
the peak. Not all
of these pairs correspond to critical sigmas, but, according to the decay
scenario described above, most of the critical sigmas certainly belong to
this set.}
\item{Adding up the momenta of the selected charged pions form the momenta
of the corresponding (fictitious) sigmas $p_{\sigma}=p_+ + p_-$.}
\item{Perform a factorial moment analysis to the momenta of the fictitious
sigmas. Due to the presence of the critical sigmas in this set, a strong
intermittency effect shows up in the sigma sector (see Fig.~5, full circles)
with the expected indices.}
\end{itemize}

\section{Numerical results}

Having developed both a Monte-Carlo generator for critical events and
an algorithm for the restoration of the critical sigma sector in an
event-by-event analysis we shall now proceed to present our numerical
results,
showing the phenomenological impact of our theoretical investigations.
Let us first specify values for our input parameters, which correspond to
the critical exponent $\delta$,
the effective coupling $g$, the transverse radius $R_{\perp}$,
the proper time scale $\tau$,
the size in rapidity $\Delta$ and the critical temperature $T_c$.
The first two parameters
are fixed, on the basis of universality class arguments, to the values of the
3-D Ising model:
$\delta \approx 5$ and $g \approx 2$ \cite{Tsyp,Wet}.
The critical temperature is taken $T_c \approx 140~MeV$.
The size in rapidity is determined so as to meet the conditions at
the RHIC collider: $\Delta \approx 11$. Finally we choose:
$R_{\perp} \approx 22~fm$ and $\tau \approx 17~fm$. It follows that
the presented results are predictions for the
event-characteristics in RHIC, provided that the critical point is there
accessible.
The above choice of $R_{\perp}$ and $g$ leads us to the value
$R_u \approx 13~fm$. This value implies the formation of 2 clusters
in the transverse momentum projection. Taking into account that $\Delta_u
\approx 0.55$ we find 10 clusters in the rapidity projection. Thus
approximately 20 critical clusters are expected to be formed in RHIC
conditions, adopting the cylindrical evolution scenario. 
Our numerical results are exhibited in a series of figures and pertain to
a large set of critical events produced
through the CMC algorithm and using parameter values appropriate for RHIC.
We investigate separately the two possible geometries, cylindrical and
spherical, for the evolution of the critical fluid.

{\it Case of cylindrical symmetry}: We produce 10000
``critical'' events and perform an event-by-event factorial moment
analysis for the sigma clusters. Our results are depicted in Figs.~1-3.
The mean multiplicity of sigmas turns out to be $\approx 90$. 
In Fig.~1a we present the
inclusive rapidity spectrum of the sigmas for the set of the 10000 events.
The uniform central plateau indicates the cancelling of the
fluctuations in the inclusive distribution.
In Fig.~1b we show the inclusive distribution of the sigmas in $p_T$.
\begin{figure}[htb]
\includegraphics[width=30pc,height=40pc]{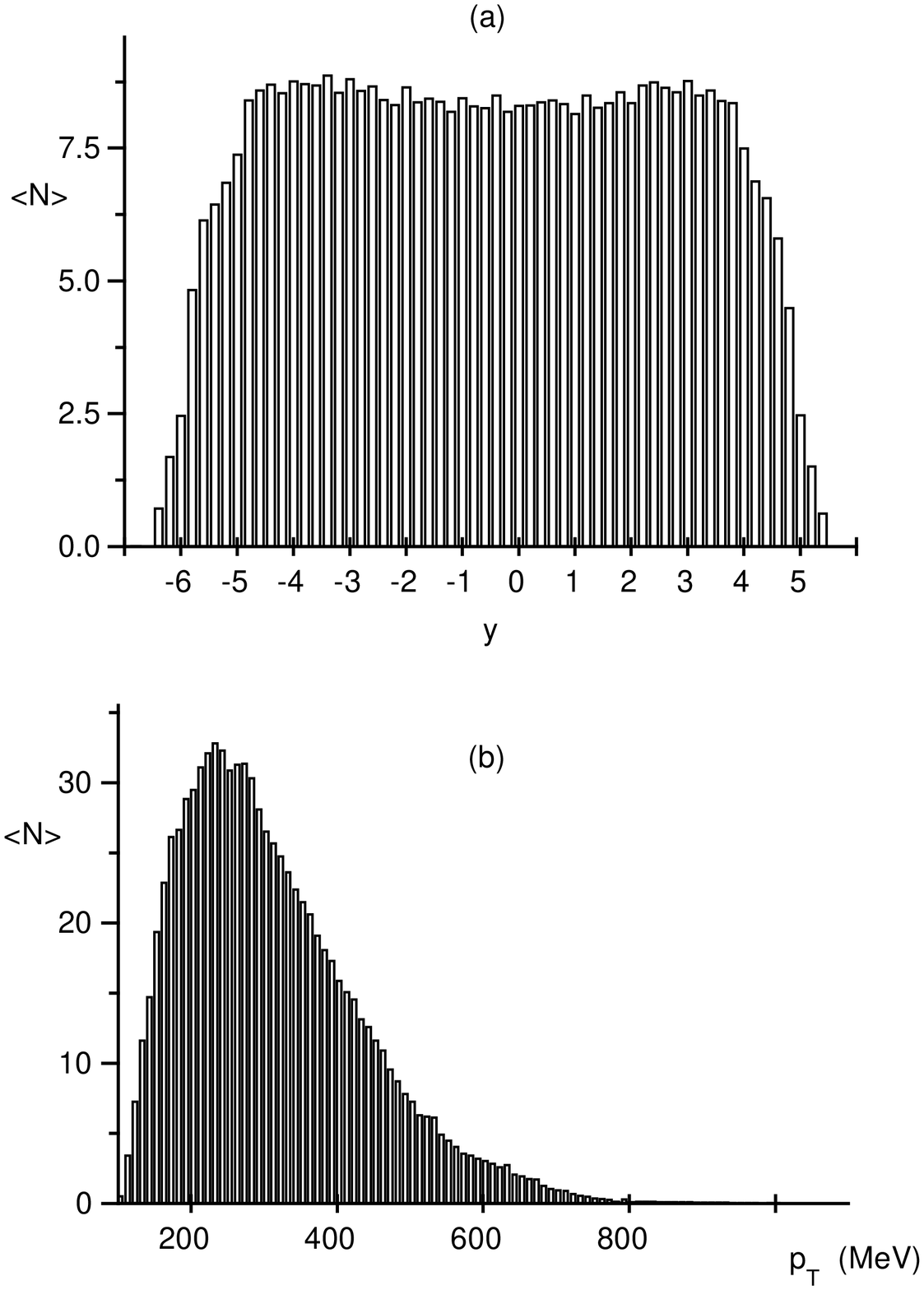}
\caption{
(a) The rapidity distribution of the $\sigma$-particles
in 10000 Monte-Carlo (MC) generated critical events (cylindrical geometry
is assumed to be valid).
(b) The corresponding inclusive transverse momentum distribution.
}
\label{fig:1}
\end{figure}

In Figs.~2a,b results of the event-by-event factorial moment
analysis in rapidity space and in transverse momentum space for the whole set of the
10000 ``critical'' events are displayed. In
Fig.~2a we present the distribution of the slopes in a linear fit for
$\ln(F_2^{(1)}(M))$ as a function of $\ln(M)$
($M$ beeing the number of bins in
rapidity) for all events. We observe a clear maximum near $s^{(1)}_2 \approx
0.27$  and a mean variance $\delta s^{(1)}_2 \approx 0.05$. This picture is
consistent
with the formation, in most events, of a complex structure with fractal
dimension $d^{(1)}_F \approx 0.73$ which is close to the mass dimension
$d^{(1)}_F = \frac{2}{3}$ obtained for a single sigma cluster (see
eq.(\ref{eq:plaws1})) in rapidity.
In Fig.~2b the analogous analysis is made for the distribution of the sigmas
in transverse momentum space.
The corresponding histogram gives the
distribution of $s^{(2)}_2$ for the slopes determined after a linear fit
to the logarithm of the two dimensional moment $F^{(2)}_2(M)$ as a
function
of the logarithm of $M$. We see, once more, a peaked
distribution with
a maximum at about $s^{(2)}_2 \approx 1.23$. This distribution is slightly
broader than the corresponding distribution in rapidity. The mean
value ($1.23$) indicates a fractal structure with dimension
$d^{(2)}_F \approx 0.77$ close to the mass dimension of a
single cluster in $p_T$ ($d^{(2)}_F = \frac{2}{3}$). 
For completeness we mention that for the
azimuth angle in transverse momentum space we get a totally uniform inclusive 
distribution, as one would expect.
\begin{figure}[htb]
\includegraphics[width=30pc,height=40pc]{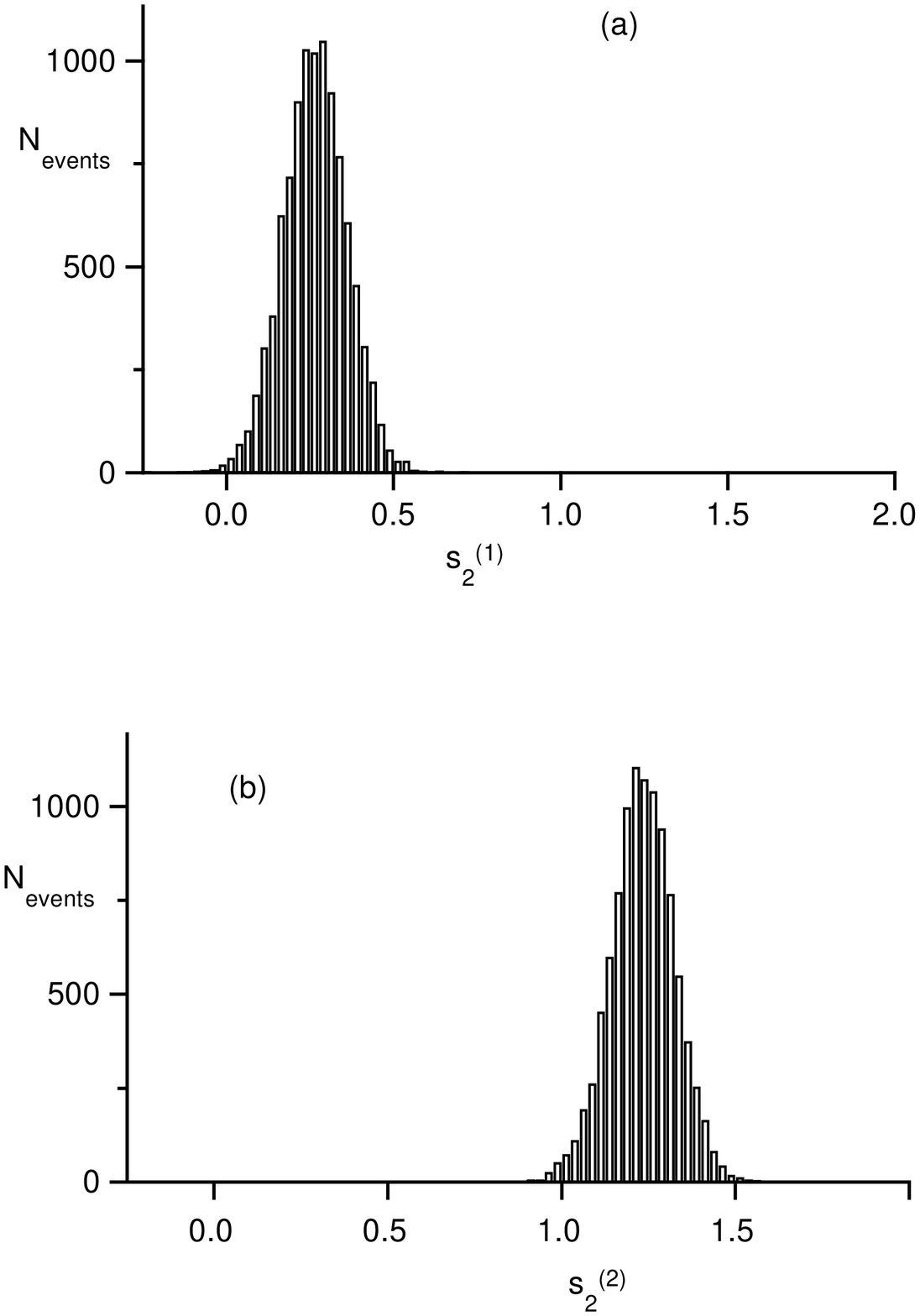}
\caption{
(a) The distribution of the slope $s^{(1)}_2$ resulting 
from a linear fit of
$\ln(F_2^{(1)})$ vs. $\ln(M)$ in an event-by-event factorial moment analysis
of the $\sigma$-particles in rapidity for the 10000 MC events
($M$ is the number of bins).
(b) The distribution of the slope $s^{(2)}_2$ obtained in the analogous analysis
in transverse momentum space for the same set of events as in Fig.~2a.
}
\label{fig:2}
\end{figure}

Putting pions into the picture we give, in Figs.~3a,b, the results of
the event-by-event factorial
moment analysis (a) in rapidity and (b) in $p_T$ of the negative pion
momentum distribution. To determine the decay of sigmas into pions
according to the distribution $P(M_{\sigma})$ we use the parameter values
$M_{\sigma}\approx 285~MeV$ and $\delta m_{\sigma} \approx 5~MeV$.
As pion mass we have used $m_{\pi}=140~MeV$. For the
mean value of $s^{(1)}_2$ in the rapidity space analysis of $\pi^-$ we find
$<s^{(1)}_2> \approx 0.07$ (see Fig.~3a) while for the case of the
transverse momentum we
find $<s_2^{(2)}> \approx 0.17$ (see Fig.~3b). These values should be compared
with the values $<s_2^{(1)}> \approx 0.27$ and $<s_2^{(2)}> \approx 1.23$
respectively, found in the event-by-event analysis of the sigmas. 
We see a clear tendency for suppression of the dynamical fluctuations in
favour of the statistical ones in the pionic sector.
\begin{figure}[htb]
\includegraphics[width=30pc,height=40pc]{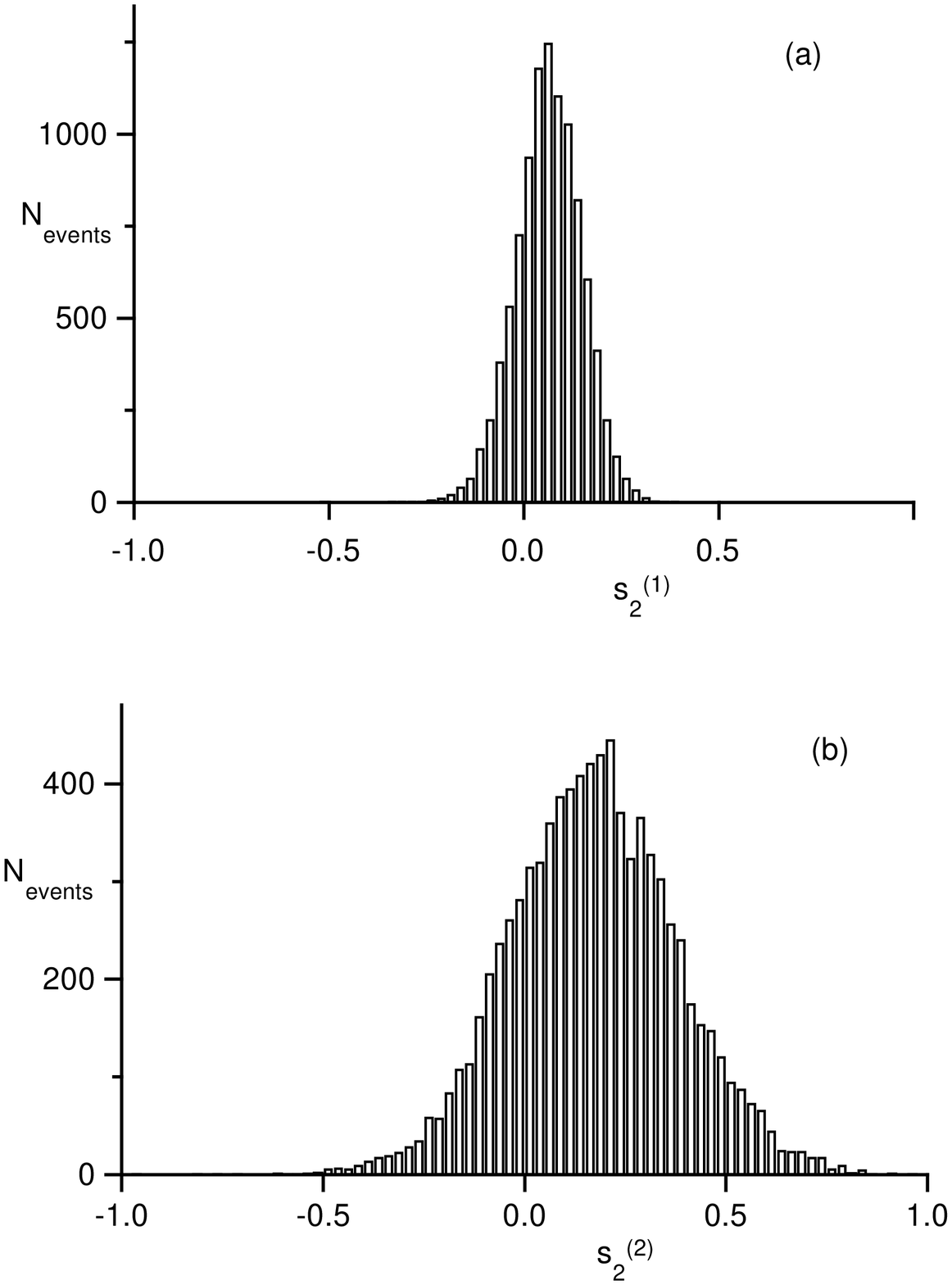}
\caption{
(a) The distribution of the slope $s^{(1)}_2$ resulting from a linear 
fit of $\ln(F_2^{(1)})$ vs. $\ln(M)$ in an event-by-event factorial moment
analysis
of the negative pions in rapidity for the 10000 MC events (the same as
in Figs.~1,2).
(b) The distribution of the slope $s_2^{(2)}$ resulting from a linear fit of
$\ln(F_2^{(2)})$ vs. $\ln(M)$ in a 2-D event-by-event factorial moment
analysis of the negative pions in transverse momentum space.
}
\label{fig:3}
\end{figure}

To recover the dynamical fluctuations one has to apply the
sigma-recontruction algorithm described in the previous paragraph.
To investigate the suppression and the restoration of the critical
fluctuations we compute the dependence of
$<s_2^{(2)}>$ for the reconstructed sigmas on the quantity
$\delta m_{\sigma}$, keeping a fixed
value $M_{\sigma} = 290~MeV$. The analysis is performed for classes (constant
$\delta m_{\sigma}$)
with 500 events each. Our results are presented in Fig.~4 (full squares).
In the same plot we show the best fit using the Boltzmann function $f(x)=
\frac{A_1 -A_2}{1+e^{(x-x_o)/d}}+A_2$ (dashed line). The fit parameters
converge to the values: $A_1=2.12$, $A_2=0.35$, $x_o=0.08$ and $d=1.1$.
The nonvanishing value of $A_2$ shows that, even if $\delta m_{\sigma}$
becomes very large, a clear intermittency signature ($<s_2^{(2)}>~\geq~0.35$)
is present in the reconstructed events.
Of course the asymptotic value of $<s_2^{(2)}>$ depends on $M_{\sigma}$ and
decreases rapidly as $M_{\sigma}$ is increased.
\begin{figure}[htb]
\includegraphics[width=30pc,height=40pc]{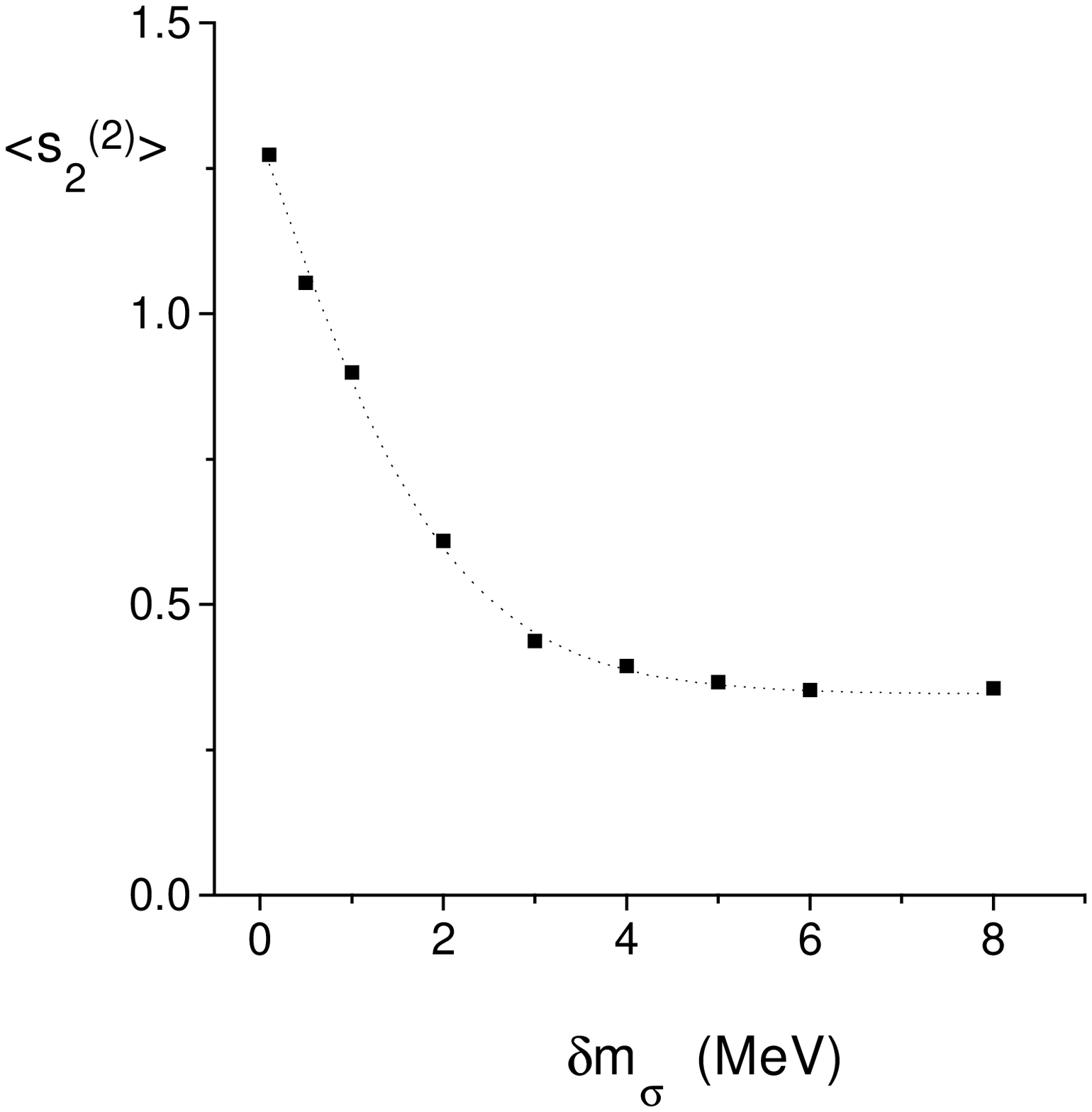}
\caption{
The dependence of $<s^{(2)}_2>$ on $\delta m_{\sigma}$
(full squares) calculated with sets each consisting
of 500 critical events generated with the cylindrically symmetric algorithm.
$M_{\sigma}$ is taken $290~MeV$.
A fit with the Boltzmann function (dashed line) is also shown.
}
\label{fig:4}
\end{figure}

To illustrate this effect more transparently we present in Fig.~5 the
second factorial moment in transverse momentum space for a typical critical
event
taking $M_{\sigma}=290~MeV$, $\delta m_{\sigma}=5~MeV$
and $m_{\pi}=140~MeV$. We see that although the fluctuations in $\pi^-$
(open up triangles) are strongly suppressed they are practically revealed
in the
sector of the reconstructed sigmas (full circles).
\begin{figure}[htb]
\includegraphics[width=30pc,height=40pc]{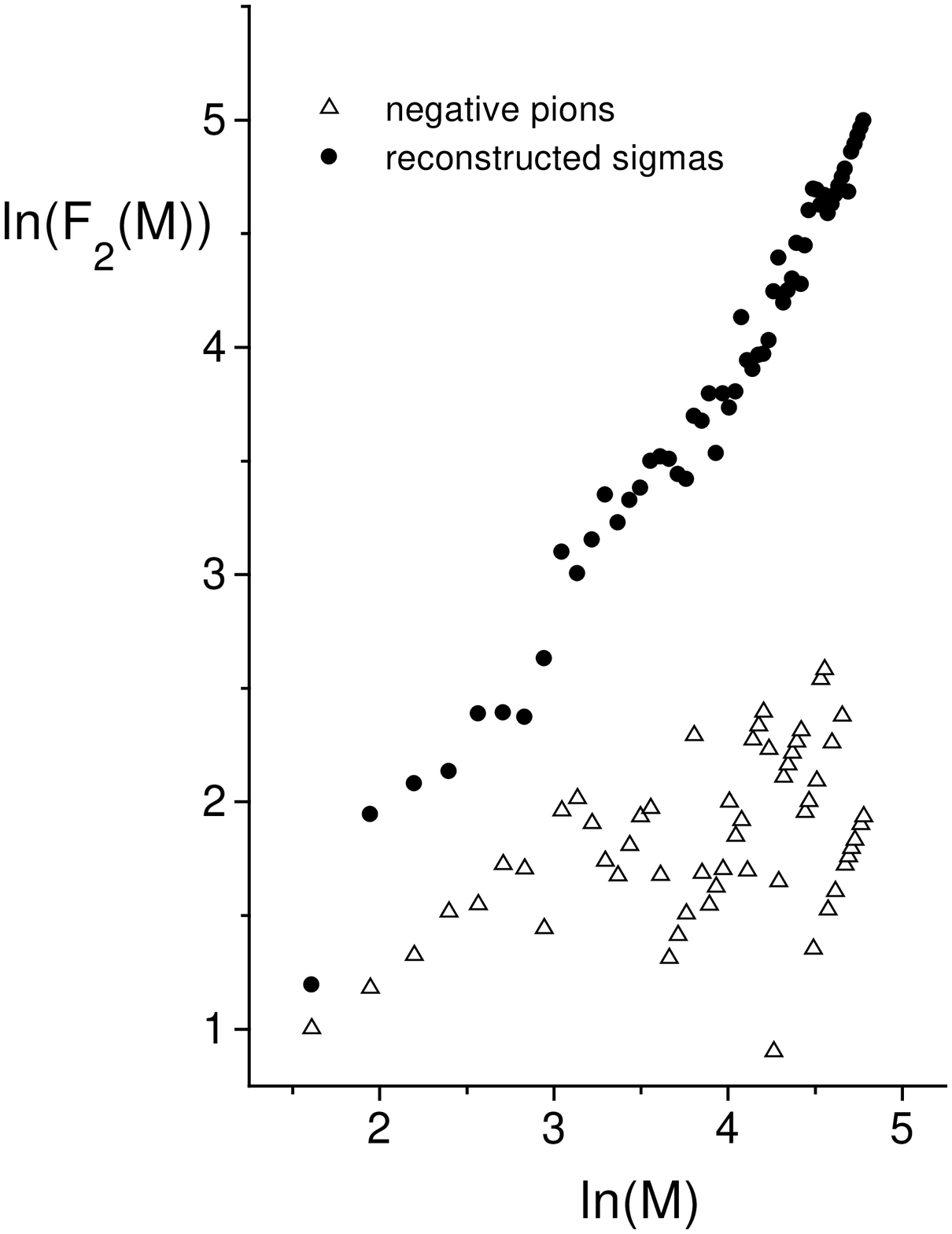}
\caption{
The second factorial moment in transverse
momentum space for a critical event (negative pions)
obtained through the cylindrically symmetric CMC generator and corresponding
to $M_{\sigma}=290~MeV$, $\delta m_{\sigma}=5~MeV$ (open up triangles).
The corresponding moment in the sigma-sector obtained
through the above described reconstruction algorithm is also shown
(full circles).
The restoration of the critical fluctuations is clearly displayed.
}
\label{fig:5}
\end{figure}

{\it Spherically symmetric evolution}:
In Figs.~6a,b we give the rapidity (a) and transverse momentum (b) 
inclusive distributions for the sigmas in a set of 10000 Monte-Carlo
generated events.
The distribution in rapidity is very interesting. The formation of a
single source (cluster) of sigmas in the collision process leads to an almost
Gaussian shape which is very clearly distinguished from the plateau emerging
in the case of many sources (cylindrical symmetry). Such differences in
the rapidity distribution are also observed between $p-p$ and heavy-ion
collision experiments far from the critical point. Our explanation 
is that the multiparticle
production in $p-p$ is dominated by the
formation of a single source while in heavy-ion collisions there are many
sources involved (supporting the scenario of longitudinal expansion).
\begin{figure}[htb]
\includegraphics[width=30pc,height=40pc]{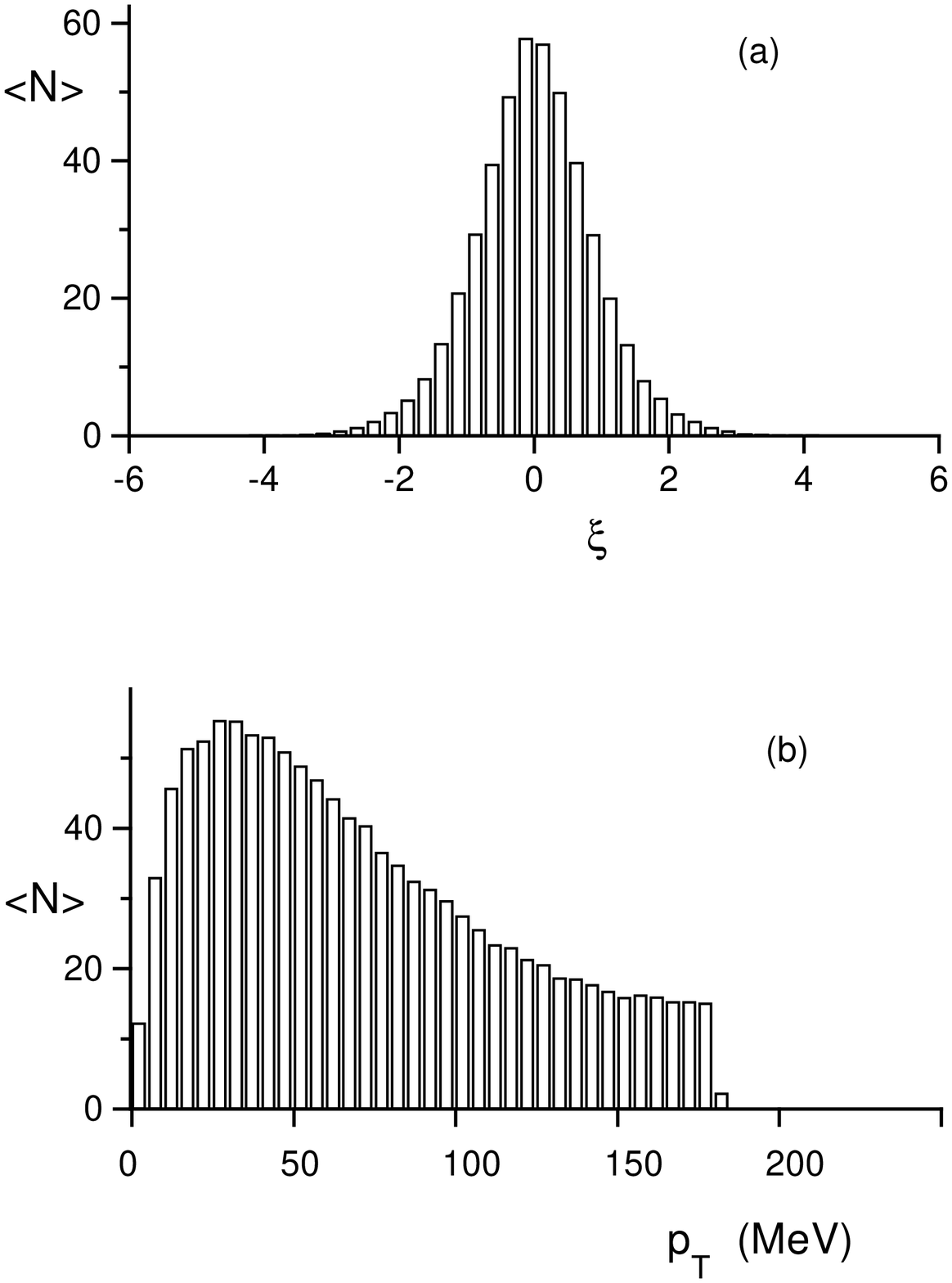}
\caption{
Inclusive distributions in (a) rapidity and (b) magnitude of the
transverse momentum of the $\sigma$-particles for 10000 Monte-Carlo generated critical
events assuming spherical symmetry.
}
\label{fig:6}
\end{figure}

Finally, in Figs. 7a,b we present the
factorial moment analysis in the sigma sector for the set of the 10000 MC
events presented in
Figs. 6a,b.
We find $<s^{(1)}_2>=0.14$ for the rapidity distribution and
$<s^{(2)}_2>=0.99$ for the
transverse momentum distribution. A suppression of the fluctuations in the
rapidity space is clearly observed. This is a peciularity of the spherical
geometry as discussed in the previous section.
\begin{figure}[htb]
\includegraphics[width=30pc,height=40pc]{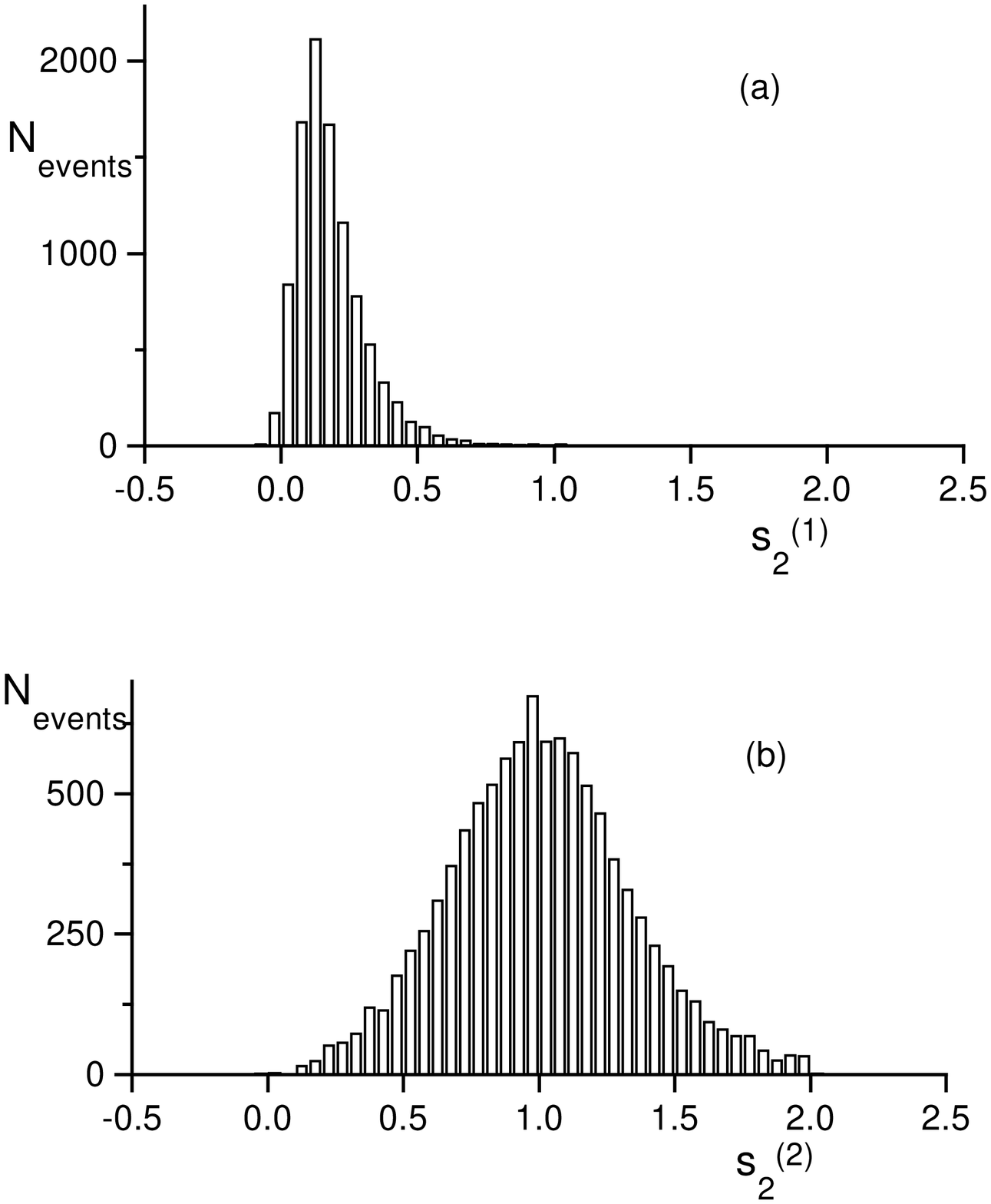}
\caption{
(a) The distribution of the slope $s^{(1)}_2$ resulting from a linear 
fit of $\ln(F_2^{(1)})$ vs. $\ln(M)$ in an event-by-event factorial moment
analysis of the sigmas in rapidity for 10000 MC events with spherical
evolution.
(b) The distribution of the slope $s_2^{(2)}$ resulting from a linear fit of
$\ln(F_2^{(2)})$ vs. $\ln(M)$ in a 2-D event-by-event factorial moment
analysis of the sigmas in transverse momentum space for the same set of
events as in (a).
}
\label{fig:7}
\end{figure}

In summary we have shown that even if the sigmas, which are expected to be
formed in a heavy ion collision experiment, as the critical point is
approached,
decay well above the two-pion threshold, it is possible, by investigating the
momentum distribution of the produced charged pions, to (partially)
reconstruct the critical sigma sector and to restore the corresponding
critical fluctuations.

\section{Concluding remarks}
 
In this work we have explored the idea that pion production in heavy ion collisions may
have a component whose source can be asrcibed to a critical QCD phase.
Our working hypothesis
has been that density fluctuations near the QCD critical point
should leave their imprints on the observed
pion spectrum. Encouraged from the accumulating evidence,
through microscopically-based
theoretical investigations, lending support to a phase diagram in which
critical QCD behaviour does occur we focused our efforts on establishing a
linkage between critical fluctuations and intermittency patterns in pion production.
Such a proposal pertains, of course, to one of two components
characterizing the pion spectrum in $A+A$ collisions,
as conventional production mechanisms are also expected to take place.

The methodology employed in this work centered itself on the statistical
mechanical aspects of the problem. We have chosen to rely on the strategy
of critical
cluster formation which seems to be the most advantageous one given
the finite size aspects of the system under study. In the context of
the currently accepted theoretical proposal that the universality class
of the critical QCD system is represented by the 3-D Ising model, we have
both established connections with pion production ascribed to the
aforementioned unconvential component and explored its observability in
terms of specific
filters associated with a reasonably wide window for the $\sigma$-mass.
Employing Monte-Carlo type numerical methodologies we have arrived at
specific {\it testable} predictions which can be checked against forthcoming
data in heavy ion experiments.
In practical terms our present undertaking can be viewed as having a dual
prospective. On the one hand, it
aims at isolating a specific type of signal in the constitution of the final
states
with an unconventional origin which would give evidence in favor of an
antecedent QGP phase. On the other, it could serve as a guide towards the
identification of the source (critical point?) responsible for long range
correlations in the QCD phase diagram \cite{Steph}.

The fact that the pions constitute the focal point of our approach is also
significant from a purely phenomenological point of view addressing
itself directly to the study of the constitution of the multiparticle
spectrum (particle multiplicities and particle ratios) produced in very
high energy collisions, especially $A+A$. In this connection, 
thermodynamical analyses from the hadronic perspective which base themselves
on thermal equilibration assumptions for the produced multiparticle
system are of special interest. Pioneering studies in this direction
\cite{HaGas} have assumed an ideal
hadron gas model and have produced satisfactory results, lending credit to
the thermal equilibration premise, but are in no position to evaluate
whether or not the source of the resulting multiparticle system is located
inside or outside (QGP?) the hadronic domain.

Recently, the same type of
analysis has been conducted \cite{Kapo} within the framework of Hagedorn's
Statistical Bootstrap Model (SBM) \cite{Boot}
once the latter was extended so as to
include the strangeness quantum number.
This approach not only incorporates, via a self-consistent logic, interactions
into the resulting thermodynamical system but also sets limit for the
hadronic phase of matter. According to the results of \cite{Kapo} there is an
accumulation, in the plane of thermodynamical variables specified by
temperature and
light quark chemical potential, of points from S+S, S+Ag and (preliminary)
Pb+Pb experimental data from SPS at CERN on either side of the curve that
separates the hadronic world from a different 
phase of matter. Matching such results from the QCD side certainly
specifies a task worthwhile pursuing (suggesting a first order transition).

Looking ahead, we assess as the most pressing question in relation to our
proposed scheme to be the following. If, according to the present theoretical
evidence, the second order phase transition achieved by the QGP system in the
temperature-baryon chemical potential plane is a critical point marking the
end of a first order transition curve, are there realistic prospects of observing 
the component contributing to pion production presently investigated? Clearly,
the answer depends as to whether one can define a reasonably `wide' window around
such a critical point within which visible implications can be formulated. This
constitutes a problem of interest which is currently being investigated.
At the same time,
we believe that the methodology developed in this paper has a wider impact
as it can be easily adopted to any future scenario concerning the, still open, issue
of QCD universality classes implicated as being realized in $A+A$ collisions.

{}

\end{document}